\documentclass[aps,prl,reprint,superscriptaddress]{revtex4-1}

\usepackage{graphicx} 
\usepackage{graphics} 
\usepackage{amsfonts} %
\usepackage{color} 

\begin{document}

\setlength{\pdfpagewidth}{8.5in}
\setlength{\pdfpageheight}{11in}

\title{Observation of EIT-enhanced cross-phase modulation in the short-pulse regime}

\author{Greg \surname{Dmochowski}}
 \email{dmochow@physics.utoronto.ca}
\author{Amir \surname{Feizpour}}
\author{Matin \surname{Hallaji}}
\author{Chao \surname{Zhuang}}
\affiliation{Centre for Quantum Information and Quantum Control, and Institute for Optical Sciences, Department
of Physics, University of Toronto, 60 St George Street, Toronto, Ontario,
Canada M5S 1A7}
\author{Alex \surname{Hayat}}
\author{Aephraim M. \surname{Steinberg}}
\affiliation{Centre for Quantum Information and Quantum Control, and Institute for Optical Sciences, Department
of Physics, University of Toronto, 60 St George Street, Toronto, Ontario,
Canada M5S 1A7}
\affiliation{Canadian Institute for Advanced Research, 180 Dundas St. W., Toronto, Ontario, Canada M5G 1Z8}

\begin{abstract}

We present an experiment using a sample of laser-cooled Rb atoms to show that cross-phase modulation schemes continue to benefit from electromagnetically-induced transparency (EIT) even as the transparency window is made narrower than the signal bandwidth (i.e., for signal pulses much shorter than the response time of the EIT system).
Addressing concerns that narrow EIT windows might not prove useful for such applications, we show that while the peak phase shift saturates in this regime, it does not drop, and the time-integrated effect continues to scale inversely with EIT window width.
This integrated phase shift is an important figure of merit for tasks such as the detection of single-photon-induced cross phase shifts.

\end{abstract}

\maketitle

The interaction between individual photons is notoriously weak.
This is an obstacle, for instance, to optical quantum computing; while the lack of interactions is helpful for coherent communications, it leaves the construction of deterministic logic gates between photonic qubits a difficult challenge.  

Non-linear optical properties of matter have been used to mediate an effective interaction between two optical fields.
For example, an intensity-dependent refractive index provided by an atomic medium can lead to a conditional phase shift written on one optical `probe' field in the presence of a second `signal' field.  
This effective interaction serves as the basis for a two-photon gate proposal, which requires a $\pi$ rad phase shift to be acquired by the probe field \cite{fredkinGate}.  
The difficulty of achieving such a large phase shift has prompted the proposal of a new approach that relies merely on the single-shot detectability of any observable cross-phase shift, not necessarily $\pi$ rad \cite{munro2005weak}.  
However, even the non-linearities required for this `weak non-linearity' scheme are many orders of magnitude larger than those achievable in typical situations.

A number of approaches have been proposed for greatly enhancing the strength of cross-phase modulation (XPM), including cavity QED \cite{turchette}, unconventional media such as photonic crystal fibers \cite{matsuda2009observation} and hollow-core fibers filled with alkali gas \cite{venkataraman2013phase}, and finally, novel physical effects such as electromagnetically induced transparency (EIT) \cite{EIT1, fleischhauer2005electromagnetically}; this Letter addresses the latter. 

The original `N-scheme' proposed by Schmidt and \.{I}mamo\u{g}lu \cite{schmidt1996giant} made use of a single EIT window, while later works investigated double-Lambda systems \cite{doubleEITtheory, doubleEITexperiment} to overcome limitations arising from the group-velocity mismatch between signal and probe pulses.
Many variations on these multi-level schemes have been introduced \cite{gateWithTripod, doubleEITtripod, gateWithInvY, Mscheme, sixLevelDoubleEIT, tripleEIT} and the ability to store light using EIT has even led to cross-phase modulation between stored pulses of light \cite{storedLightXPM}. 
In the past 3 years, remarkably high non-linearities have been observed relying on Rydberg blockade in the presence of EIT \cite{peyronel2012quantum, firstenberg2013attractive, Chen16082013, PhysRevLett.112.073901}, presaging potentially huge cross-phase shifts.





The enhancement provided by EIT schemes arises from the spectral narrowness of the transparency window, which results in a steep refractive index profile for the probe field. 
The presence of the signal field introduces an AC Stark shift, which effectively detunes the probe field out of resonance, causing it to acquire a phase shift proportional to the slope of this refractive index profile. 
Meanwhile, the AC Stark shift is intensity-dependent: a larger shift will be produced by compressing the signal energy into a temporally shorter pulse.  
Presumably, then, the largest non-linear effect is achieved when spectrally narrow transparency windows are perturbed by short, intense signal pulses. 
However, reducing the spectral width of the transparency window, $\Delta_{EIT}$, implies a slower response time for the EIT medium. 
This suggests that a fundamental limitation may exist for EIT-enhanced XPM schemes; once the inverse EIT bandwidth, $1/\Delta_{EIT}$, exceeds the 
temporal width of the signal pulse, $\tau_s$ (i.e.once the signal pulse is shorter in time than the response time of the medium), it appears as though the EIT medium could not respond quickly enough to provide a practical benefit. 
Theoretical \cite{pack2006transients, sinclair2009time} and experimental \cite{pack2007transients} investigations of step-function signal fields have, indeed, found that narrower windows, while providing a larger steady state phase shift, require more time to reach this steady state.
The authors went on to conclude that such a slow response time may be a limitation in the case of pulsed signal fields.
However, there has been no study of the transient behaviour of EIT-enhanced XPM in this practically more relevant scenario where the signal is a (broadband) pulse.
It has thus far been unclear whether the advantages of the original theoretical proposal \cite{schmidt1996giant} (which treated only the single-mode problem) hold in practice. 

Here we show experimentally that narrow transparency windows continue to provide a benefit for XPM schemes even when $\tau_s \ll 1/\Delta_{EIT}$. 
The slow dynamics reported for step-responses \cite{pack2006transients, sinclair2009time, pack2007transients} are actually at the root of the enhancement offered by EIT in the regimes of most practical interest, namely, narrow transparency windows perturbed by short, intense signal pulses.  
While the peak phase shift saturates when $\tau_s \leq 1/\Delta_{EIT}$, it does not decrease as $\Delta_{EIT} \rightarrow 0$, and the narrow EIT bandwidth serves to prolong the effect of the short signal pulse.
As a result, we find that in the short-pulse regime, the integrated phase shift (though not the peak phase shift) continues to scale linearly with $1/\Delta_{EIT}$.
For weak non-linearity schemes such as \cite{munro2005weak}, which rely on the mere detectability of a single-shot cross-phase shift (not necessarily $\pi$ rad), this amounts to a significant advantage over non-EIT schemes.
This work constitutes the first study of EIT-enhanced XPM in this experimentally relevant regime of spectrally narrow transparency windows perturbed by broadband signal pulses.

\begin{figure}[htb]
   \centering
   \includegraphics[scale=0.35,bb=10 10 700 390]{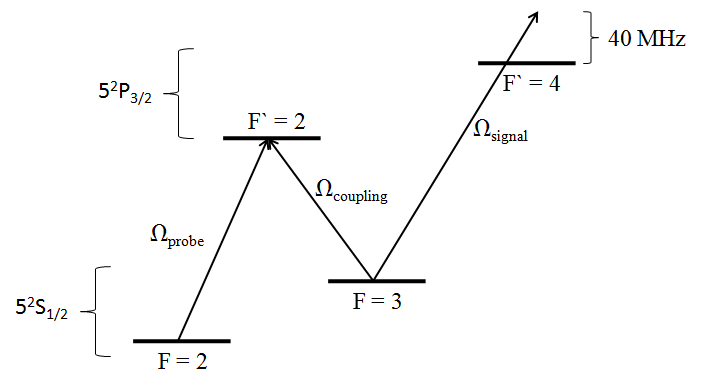}
   \caption{Level scheme used to implement EIT-enhanced XPM in $^{85}$Rb atoms.  A probe beam ($\Omega_{\text{probe}}$) addressing the $5^2$S$_{1/2} \rightarrow 5^2$P$_{3/2}$ transition experiences EIT due to a strong coupling beam ($\Omega_{\text{coupling}}$), while a detuned signal pulse ($\Omega_{\text{signal}}$) serves to AC Stark shift the F = 3 ground state, producing a cross-phase shift on the probe field.}
   \label{Nscheme}
\end{figure}

We study the effects of EIT on XPM using an ensemble of $^{85}$Rb atoms cooled and trapped in a magneto-optical trap (MOT). 
The level scheme in our experiment is shown in figure \ref{Nscheme} and makes use of the two $5^2$S$_{1/2}$ hyperfine ground states along with the $5^2$S$_{3/2}$ $F ' = 2$ excited state to form a three-level $\Lambda$-system. 
When the probe and coupling fields satisfy the two-photon resonance condition (i.e. when their respective detunings are equal), the atomic medium is rendered transparent to the probe beam. 
The degree of transparency (as well as the refractive index) experienced by the probe beam depends strongly on its detuning from this two-photon resonance.
Under EIT conditions, pulses of signal light blue-detuned by $40$ MHz from the $F = 3 \rightarrow F ' = 4$ transition are sent through the medium, temporarily AC Stark-shifting the $F = 3$ hyperfine ground state.
This energy-shift of the ground state changes the two-photon detuning, imprinting a phase shift on the probe field which, in the steady-state, is inversely proportional to the EIT window width. 

The phase shift is measured using a beat-note interferometry \cite{beatNote} technique (the frequency-domain analog of a spatial interferometer).  
A `reference' beam, co-propagating with the probe field but blue-detuned by $100$MHz, is also sent into the medium, serving as a phase reference.  
This detuning is large enough that the phase shift imprinted on the reference beam due to the signal-induced AC Stark shift is negligible.  
The phase shift of the on-resonance probe field manifests itself as a phase shift of the resulting beat signal, which is read out by demodulation at $100$MHz. 
The coupling and signal fields also co-propagate with these two probing fields, and all the beams are focused down to a beam waist of $\approx 50\mu$m centered on the $2$mm diameter atomic cloud.
After an initial $10$ms cooling time, the magnetic field gradient and trapping lasers are turned off for $1$ms, leaving only the CW probing and coupling fields on while pulses of signal light are sent into the medium.
This cycle is repeated and every data point presented below corresponds to an average of several thousand pulses.
The cloud of atoms has a density of $\approx 10^{10}cm^{-3}$, and is cooled down to $\approx 50 \mu$K, with an optical density of $3$ as seen by the on-resonance probe beam. 
We observed $\approx 15\%$ atom number fluctuations from shot to shot throughout the experiment.
The coupling and signal fields have the same polarization, which is orthogonal to the polarization of the probing fields, allowing for the detection of the probing fields alone.

\begin{figure}[ht]
  \includegraphics[width = \columnwidth]{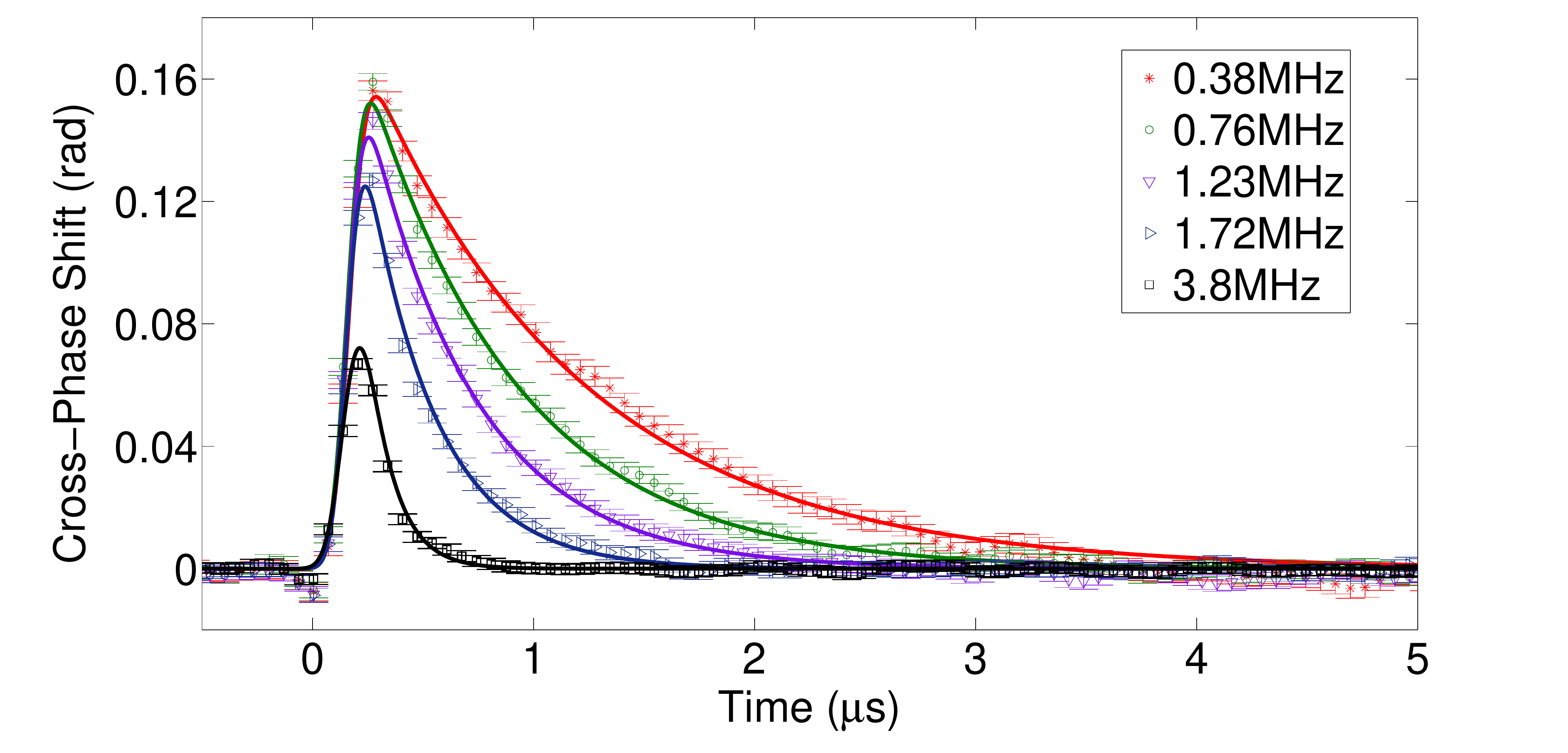}
  \caption{Temporal profile of XPM for a variety of transparency window widths (FWHM shown in the legend).  A $40$ns (RMS) Gaussian signal pulse was incident at $t=0$. The phase shift of the probe field rises with a timescale given by the signal pulse duration, irrespective of $\Delta_{EIT}$.  The decay, on the other hand, is governed by the larger of $\tau_s$ and $1/\Delta_{EIT}$. For $1/\Delta_{EIT} > \tau_s$, the phase returns to its original steady-state value at a rate given by the transparency window width. In addition, the peak phase shift saturates as the window width is narrowed. The fits to the data (solid lines) are obtained from a linear time-invariant system model of the interaction \cite{XPMtheory}, where the phase of the probe field is treated as the output of a linear system that is being driven by the signal pulse intensity.
  \label{timeTraceWindows}}
\end{figure}

\begin{figure}[t]
  \centering %
  \includegraphics[width = \columnwidth]{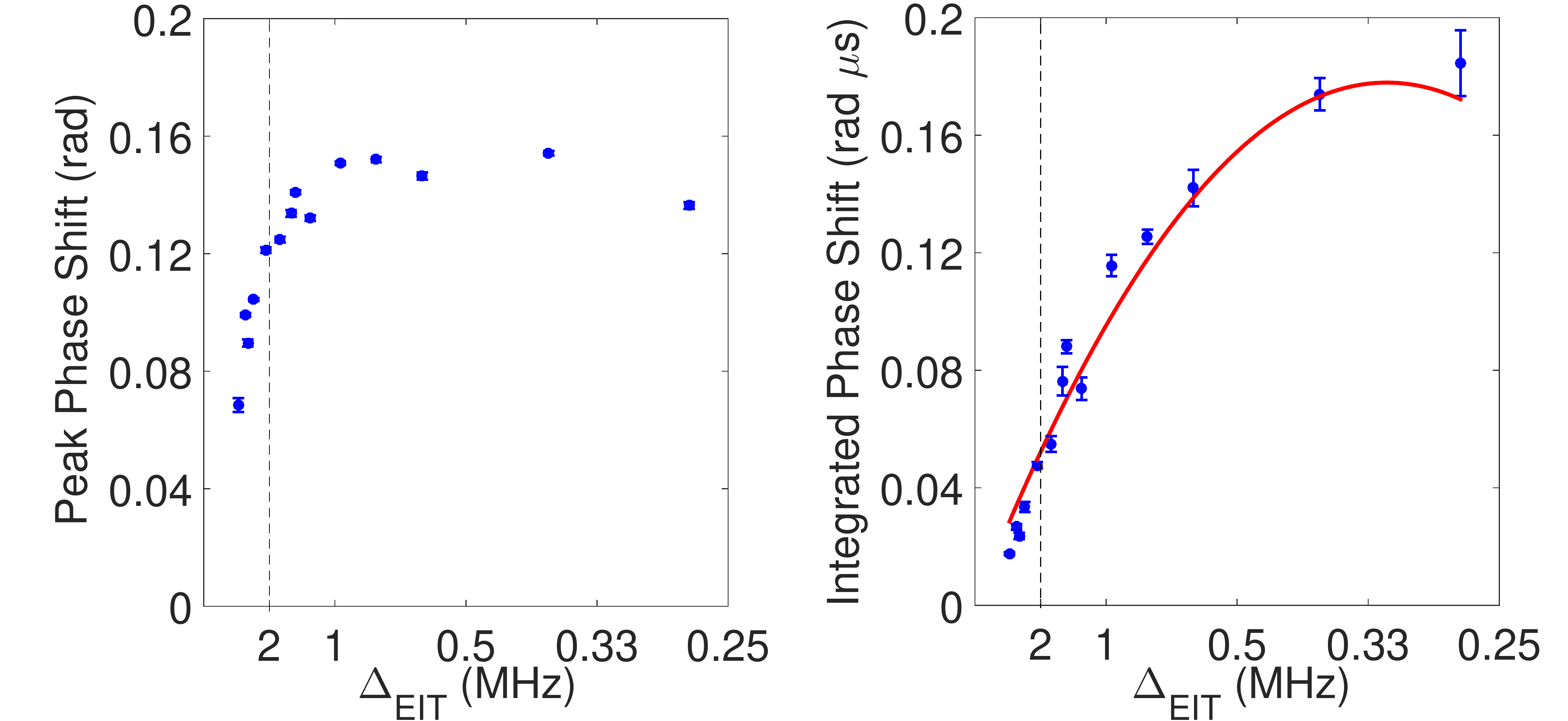}
  \caption{Peak (left) and integrated (right) phase shifts as extracted from figure \ref{timeTraceWindows} plotted against EIT window width (note inverted scale). Decreasing the EIT window by an order of magnitude results in an integrated phase shift that grows proportionally while the peak phase shift saturates after increasing by a factor of two. The dashed, vertical line at $2$MHz corresponds to the bandwidth of the signal pulse and shows where the EIT window width becomes spectrally narrower than the signal pulse; the peak phase shift saturates at this point while the integrated phase continues to grow. The deviation from linearity of the integrated phase shift, as well the eventual dip of the peak phase, arises when the transparency window becomes comparable to the dephasing rate of the EIT system. The red curve is a fit to the data using an expression for the integrated phase shift from a linear time-invariant system model for the interaction which includes this dephasing rate\cite{XPMtheory}.
  \label{peakAndIntegratedWindows}}
\end{figure}

Figure \ref{timeTraceWindows} shows temporal profiles of the cross-phase shift acquired by the probe beam due to a $40$ns (RMS), $0.8\mu$W peak power Gaussian signal pulse for a variety of EIT window widths. 
The widths of the transparency windows were measured separately (by scanning the frequency of the probe laser across the two-photon transition) and their FWHM are provided in the legend.
The coupling beam powers ranged from $192$nW to $5\mu$W depending on the desired EIT window width while the on- and off-resonance probe powers were kept at $8$nW and $100$nW, respectively, throughout.
Each trace in figure \ref{timeTraceWindows} is an average of $2500$ signal pulses.
It can be seen that the rise of the phase shift is independent of the EIT window, in contrast to the previously observed step-response behavior. 
While the peak phase shift saturates for small $\Delta_{EIT}$, it does not decrease. 
In addition to this, the effect of the passing signal pulse is seen to last longer as $\Delta_{EIT}$ is narrowed. 
For example, in the case of a $0.38$MHz transparency window, the $40$ns signal pulse produces a phase shift that decays with a $1/e$ time of $1.1\mu$s.
 
The solid curves in figure \ref{timeTraceWindows} are fits to the data based on a linear time-invariant (LTI) model of the light-matter interaction; in the cross-Kerr regime, we can abstract the underlying non-linearity and treat the phase of the on-resonance probe field as the output of a linear system, which is being driven by the signal field intensity.
This phase, then, is given by the convolution of the Gaussian signal pulse with the impulse response function of the EIT system, which can be well approximated by a decaying exponential with decay constant, $\tau$.
The result of the convolution yields the following expression for the time-dependence of the non-linear phase shift:

\begin{equation}
\phi(t) = \frac{\phi_0 n_{ph}}{2\tau} e^{\tau_s^2/2\tau^2} e^{-t/\tau}\left[1+ \text{erf}\left( \frac{t}{\sqrt{2}\tau_s} - \frac{\tau_s}{\sqrt{2}\tau}\right) \right]
\label{eqTemporalProfile}
\end{equation}

\noindent
where $\phi_0$ is the integrated phase shift per photon, $n_{ph}$ is the number of signal photons, and $\tau_s$ is the RMS duration of the signal pulse intensity.
In a separate work \cite{XPMtheory}, we have theoretically investigated the validity of this approach and found excellent agreement with a rigorous semi-classical treatment of the problem. 
The work presented here serves as experimental verification of such an LTI model as the fits in figure \ref{timeTraceWindows} match the experimental data well.
Both the signal pulse duration, $\tau_s$, and the time constant of the decaying exponential, $\tau$, were left as fitting parameters and the results agree very well with the known values.

%

We now consider how the magnitude of the cross-phase shift is affected by narrowing the EIT window.
Figure \ref{peakAndIntegratedWindows} plots the peak and integrated phase shifts (as extracted from figure \ref{timeTraceWindows}) versus $\Delta_{EIT}$.
While single-mode (stationary state) as well as step-response treatments yield a linear dependence of phase shift on $1/\Delta_{EIT}$, the peak phase shift is seen to saturate as $\Delta_{EIT}$ is narrowed (left to right in the figure) in the presence of a broadband signal pulse.
The vertical, dashed lines in figure \ref{peakAndIntegratedWindows} denote the $2$MHz bandwidth of the signal pulse and mark the onset of saturation in the peak phase shift.
For purposes of achieving $\pi$ rad phase shifts this clearly presents a limitation.
The integrated phase shift, on the other hand, is seen to continue its linear growth far into the regime where the peak phase shift has saturated.
Decreasing the EIT window by an order of magnitude causes the integrated phase shift to grow by nearly a factor of ten while the peak phase shift grows by only a factor of two. 
In fact, it is only the finite coherence time of our system that disrupts the linear scaling of the integrated cross-phase shift with $1/\Delta_{EIT}$ \cite{XPMtheory}.
The red curve in figure \ref{peakAndIntegratedWindows} is a fit to the data using an expression for the integrated phase shift, $\phi_0 n_{ph}$, which includes this dephasing.
Once again, we see good agreement with experiment.
The dephasing rate was measured separately to be $75$kHz and is due to lingering, inhomogeneous magnetic fields that persist during the measurement even after we have turned off the magnetic field gradient.
Ground-state dephasing serves to limit the depth of transparency, as well as to broaden the EIT window. 
It can be shown that as $\Delta_{EIT}$ nears the dephasing rate, $\gamma$, the integrated phase shift will saturate, peaking when $\Delta_{EIT} = 4\gamma$. 

Given that the peak phase shift saturates, the linear growth of the integrated phase shift is entirely due to the longer decay time provided by narrow windows.
The temporal profile of the cross-phase shift (eq. \ref{eqTemporalProfile}) depends on both the signal pulse duration, $\tau_s$, and the EIT response time, $\tau$.
In the narrow-EIT regime ($\tau \gg \tau_s$), these can be regarded as the rise and fall times, respectively, of the cross-phase shift in figure \ref{timeTraceWindows}; for short times $t$, the error function term dominates, its time scale set by $\tau_s$, whereas for longer times the decaying exponential takes over with a time constant determined by the EIT system.

\begin{figure}[t]
  \centering %
  \includegraphics[width = \columnwidth]{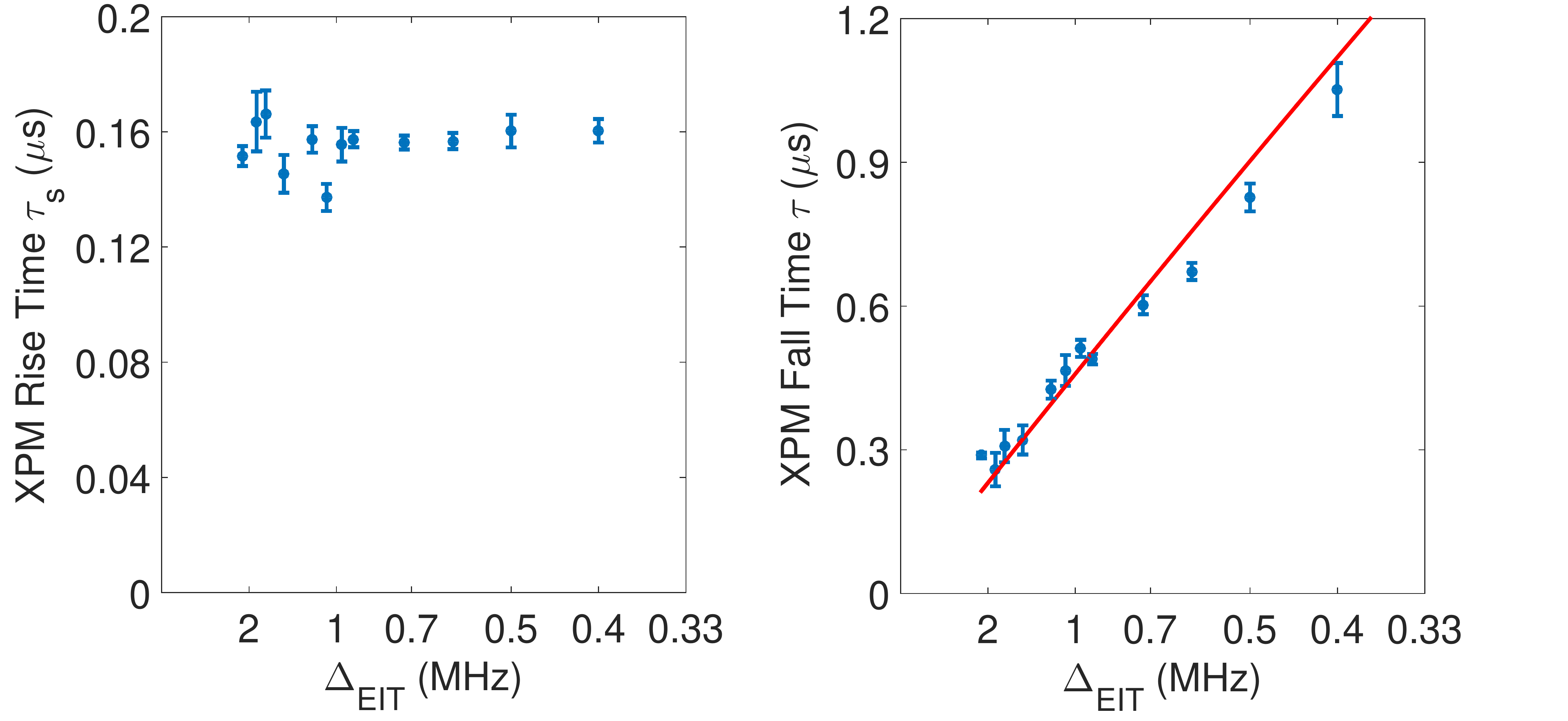}
  \caption{Rise time, $\tau_s$, (left) and fall time, $\tau$, (right) of XPM plotted against EIT window width (note inverted scale) as extracted from fitting the data in figure \ref{timeTraceWindows} to eq. \ref{eqTemporalProfile}.  As the EIT window is narrowed (left to right), the decay of the XPM is slower, while the rise time is unaffected, being given by a convolution of the measurement sampling period ($67$ns) and the signal pulse duration ($140$ns). The red line is a parameter-free theory curve, eq. \ref{eqEITresponseTime}, describing the response time of the EIT medium.
  \label{riseAndFallTimesWindows}}
\end{figure}

To study the behavior of these dynamics, we repeated the measurement shown in figure \ref{timeTraceWindows} but with lower signal intensity so as to avoid saturating the atoms; a longer signal pulse duration of $140$ns (RMS) was used along with a lower peak power ($160$nw) and an OD of $1.8$.
We fit the resulting temporal profiles, leaving $\tau_s$ and $\tau$ as free parameters.
These extracted parameters are plotted in figure \ref{riseAndFallTimesWindows} versus EIT window spectral width, $\Delta_{EIT}$.
The roughly constant rise time of approximately $160$ns arises from the smearing out of the signal pulse duration by the measurement sampling period of $67$ns. 
The fall time, on the other hand, is seen to grow as the EIT window is narrowed.
The presence of the signal pulse inside the medium disrupts the atomic coherence established by the probe and coupling fields. 
Upon passage of the signal pulse, this coherence returns to its steady-state value but on a timescale governed by the width of the EIT window (i.e. by the intensities of the probe and coupling fields). 
The lower field strengths necessary for narrow EIT windows yield a slower optical pumping rate into the dark state, and therefore, a longer time is needed to return to steady-state.
The red line in figure \ref{riseAndFallTimesWindows} is a theoretical curve (with no free parameters) for the fall-time, $\tau$, of the phase shift:

\begin{equation}
\tau = \left[1 + \frac{d_0}{4}\left(1 - \frac{2\gamma}{\Delta_{EIT}} \right)\right]\frac{2}{\Delta_{EIT}},
\label{eqEITresponseTime}
\end{equation}

\noindent
where $d_0$ is the optical density and $\gamma$ is the ground state dephasing rate of the EIT system.
This expression first appears as the exponential rise-time of cross-phase modulation in the case of a step-function signal field \cite{pack2006transients}. 
Since the impulse response is the derivative of the step-response, this expression carries over as the decay constant, $\tau$, found in the exponential term in eq. \ref{eqTemporalProfile}.
That is, we see that the decay of the cross-phase shift in the wake of broadband signal pulses ($\tau_s \ll 1/\Delta_{EIT}$) obeys the same dynamics as the step-response of XPM.
In both cases, a linear system is evolving toward a steady state - either returning to its original steady state upon passage of the signal pulse or progressing towards its new steady state in the presence of a step-signal field. 
This is why we conclude that the slow dynamics reported for the step-response of EIT-enhanced XPM are actually at the root of the enhancement for pulsed-signal XPM when $\tau_s \ll 1/\Delta_{EIT}$. 

\begin{figure}[t]
  \centering %
  \includegraphics[width = \columnwidth]{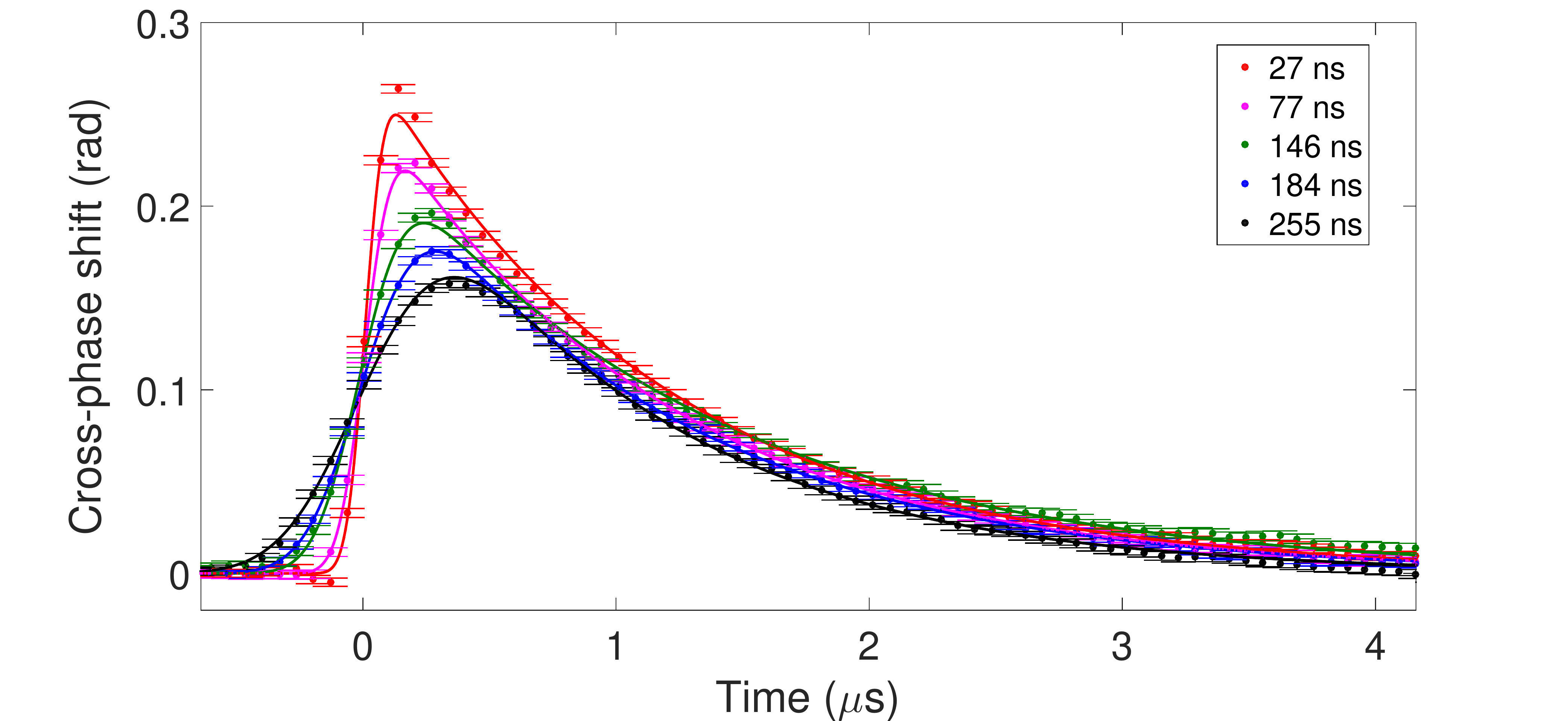}
  \caption{Temporal profile of XPM for a variety of signal pulse durations, holding the total signal energy fixed at $75$fJ (corresponding to $300,000$ photons per pulse), with an EIT window of $600$kHz. The peak of the pulses arrived at $t = 0$. While the decay time of the phase shift is constant due to the fixed EIT bandwidth, the rise time is seen to depend on the signal pulse duration. The fits to the data (solid lines) are obtained from a linear time-invariant system model of the interaction \cite{XPMtheory}.
  \label{timeTracesPulses}}
\end{figure}

Finally, to confirm that the rise time of EIT-enhanced XPM is, indeed, governed by the signal duration, we studied the effect of varying the signal pulse width.
Figure \ref{timeTracesPulses} shows temporal profiles of the cross-phase shift for different signal pulse widths, while keeping the energy of each signal pulse constant at $ 75$fJ, which corresponds to $\approx 3\times10^5$ photons.  
A $600$kHz EIT window was used for these measurements, which results in the constant decay times seen in the figure, and the on-resonance probe field experienced an OD of 3. 
The signal pulses were detuned by $40$MHz and were of Gaussian shape with $\tau_s$, the rms width of the signal, ranging from $20$ns to $255$ns.
This is reflected in the range of rise times observed in figure \ref{timeTracesPulses}, with each data point being an average of $2500$ signal pulses.  
The zero of the time axis corresponds to the arrival of the peak of each signal pulse. 

\begin{figure}[t]
  \centering %
  \includegraphics[width = \columnwidth]{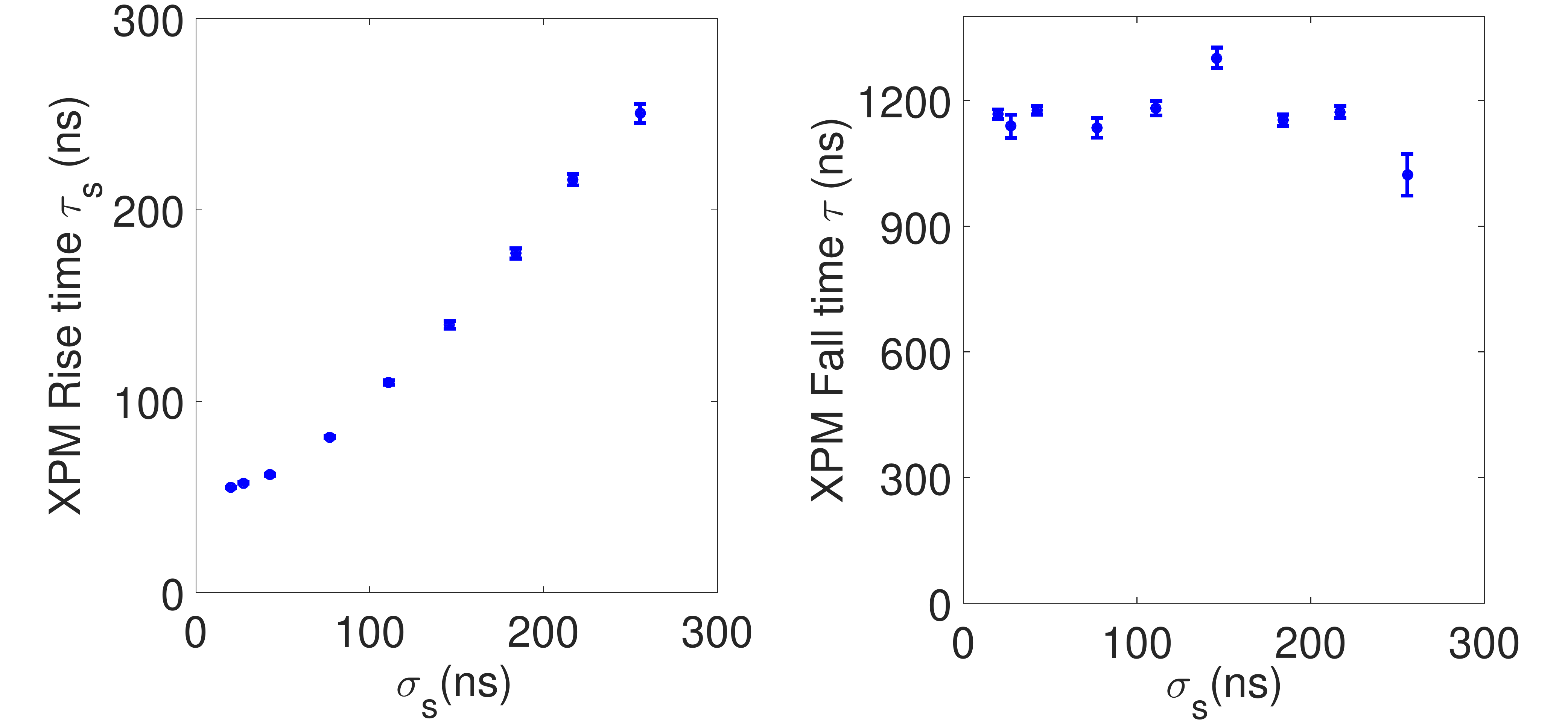}
  \caption{Rise (left) and fall (right) times of XPM extracted from Fig. \ref{timeTracesPulses} as a function of signal pulse (rms) duration, $\sigma_s$.  The rise time grows with the signal pulse duration whereas the fall time is seen to be nearly flat since $\Delta_{EIT} = 600$kHz $< 1/\sigma_s$ for the signal pulses used. The deviation from linearity of the rise time arises when the pulse duration becomes comparable to the $67$ns sampling period.
  \label{timescalesPulses}}
\end{figure}

Figure \ref{timescalesPulses} shows the extracted rise and fall times as a function of signal pulse duration.  
While the decay time of XPM appears roughly constant, the rise time is clearly seen to depend on the signal pulse duration.  
For signal pulses much wider than our $67$ns sampling period, a clear linear dependence is observed, confirming that the rise of XPM is determined by the signal pulse duration. 
For signal pulse durations comparable to the sampling period, we see a rise time that is smeared out due to the measurement. 
Importantly, while the EIT window for these measurements was $600$kHz (corresponding to a timescale of $\approx 830$ns), the rise time of XPM is seen to be several times faster, clearly not limited by $\Delta_{EIT}$.
The attentive reader may have noticed that a $600$kHz window here gave a fall time of $1200$ns while in figure \ref{riseAndFallTimesWindows}, the same window width yielded approximately half this value. 
This is due in part to the higher OD used in figure \ref{timeTracesPulses} but also to the relatively high signal power used in this data set.
We have seen that sufficiently high signal powers yield decay times longer than those predicted by eq. \ref{eqEITresponseTime}.
The dependence of the peak and integrated phase shifts on signal bandwidth are discussed in full detail in \cite{XPMtheory}.

We have experimentally demonstrated that narrow EIT windows continue to provide a benefit for cross-phase modulation schemes even when the signal pulse duration is much shorter than the inverse window width. 
Despite concerns that a narrow transparency window is necessarily accompanied by a slow response, we see that the cross-phase shifts rise on a timescale given by the signal pulse duration, which can be orders of magnitude faster than the inverse EIT window width.
While the peak phase shift saturates when the signal pulse duration becomes shorter than the inverse window width, the integrated phase shift continues to scale linearly with $1/\Delta_{EIT}$.
With respect to signal pulse duration, there is no minimum acceptable EIT window width: the integrated phase shift will continue to increase as the window is narrowed, limited only by ground state dephashing of the EIT system. 
The extended duration of the cross-phase shift afforded by narrow EIT windows allows for a longer measurement time, which can bring about a significant reduction in noise.
This enhancement is very promising for all-optical quantum computing schemes, as well as other applications relying on the non-linear effects of low-energy optical pulses.

\section*{Acknowledgement}
We would like to thank Tilman Pfau for helpful discussions and first raising the concern of using broadband signal pulses with narrow EIT windows, as well as Alan Stummer for continued technical support and John Howell for helpful discussions. This work was funded by NSERC, CIFAR, and QuantumWorks.

\bibliographystyle{apsrev4-1}
\bibliography{refs}

\end{document}